\documentclass[a4paper,11pt]{article}
\usepackage{jcappub} % for details on the use of the package, please see the JINST-author-manual
\usepackage{gensymb}
\usepackage{subcaption}
\usepackage{ulem}
\usepackage{cancel}

\usepackage{hyperref}       % for \href, \url
\usepackage{soul}
\usepackage{textcomp}       % for \textit, \textquotes
\usepackage{amsmath}        % for math symbols
\usepackage{amsfonts} 
\usepackage{lmodern} % Latin Modern fonts support bold
\usepackage{natbib}  
\usepackage[utf8]{inputenc}   % allows Unicode (degree, etc.)
\usepackage[T1]{fontenc}      % proper font encoding
\usepackage{textcomp}         % defines \degree
\usepackage[dvipsnames]{xcolor}
\usepackage{xcolor}           % enables colored text
\newcommand{\rev}[1]{\textcolor{red}{#1}}
\newcommand{\err}[2]{^{+\!#1}_{-\!#2}}

\renewcommand{\textcolor}[2]{#2}

\usepackage{url}

\usepackage{breakurl}

\title{ X-ray and Radio Analysis of Abell 1644: Constraints on Cluster Dynamics}

\author{Humaira Bashir\textsuperscript{1}, R. Kale \textsuperscript{2}, Asif Iqbal\textsuperscript{3} and Manzoor A. Malik\textsuperscript{1*}}
\affiliation{\textsuperscript{1}Department of Physics, University of Kashmir, Hazratbal, 190006, India.}
\affiliation{\textsuperscript{2}National Centre for Radio Astrophysics, Tata Institute of Fundamental Research, Pune 411007, India.}
\affiliation{\textsuperscript{3}Univ. Lille, Univ. Artois, Univ. Littoral C\^ote d'Opale, ULR 7369 -- URePSSS -- Unit\'e de Recherche Pluridisciplinaire Sport Sant\'e Soci\'et\'e,  F-59000 Lille, France}

\emailAdd{mmalik@kashmiruniversity.ac.in}

\abstract{We present the first band-2 (120--250\,MHz) uGMRT (upgraded Giant Metrewave Radio Telescope) observations of the bimodal galaxy cluster Abell\,1644 (\(z = 0.0471\)), complemented by Chandra X-ray data. While weak lensing measurements reveal a third substructure in Abell 1644, our radio analysis reveals only two compact sources coinciding with the respective brightest cluster galaxies (BCGs) of the northern (A1644N1) and southern (A1644S) substructures, seen in the X-ray observations. Radio analysis yields compact active galactic nuclei (AGN) powered sources with radio power $P_{A1644S} = \textcolor{red}{1.0}\times 10^{23} W/Hz$ and $P_{A1644N} = \textcolor{red}{7.9}\times 10^{23} W/Hz$ at 200MHz. We find no evidence of non-thermal diffuse radio emission, such as halos or relics, within the sensitivity of our band-2 image.  We measured the flux density of each radio source and performed spectral analysis. A1644N1  exhibits a synchrotron power law spectrum while A1644S shows spectral turnover suggestive of synchrotron self-absorption. Our X-ray analysis confirms the presence of a cold front east of the A1644S subcluster core. The temperature map further reveals a previously unreported asymmetry, with a hot intracluster medium (ICM) region to the east of A1644S and cooler gas to the west, likely representing residual signatures of earlier merger activity. Together, these features indicate that Abell 1644 preserves clear imprints of its merger history through long-lived sloshing motions, while the absence of diffuse radio emission suggests that the past merger was relatively minor not injecting enough turbulence for large scale reacceleration or the  cluster is approaching a late stage of ICM relaxation.}


\begin{document}
\maketitle
\flushbottom

\section{Introduction} 
Galaxy clusters are the largest gravitationally bound structures in the universe, with masses up to $ 10^{15} M_{\odot}$. These clusters form the nodes of the universe’s large-scale, filament-like structure through a bottom-up evolutionary process. Although clusters can contain thousands of galaxies, these galaxies only account for about $5\%$   of the total mass. The bulk of the baryonic matter, roughly $15\%$, is contained in a hot, ionized intracluster medium (ICM) with temperatures between $10^{7} - 10^{8}$ K, held together by the cluster’s gravitational forces. The remaining mass, $\sim 80\%$, is thought to be dark matter \citep{white,Pratt}.

The ICM, a diffuse plasma composed of thermal particles with electron densities of approximately $10^{-3} -10^{-4} cm^{-3}$, also contains relativistic particles and magnetic fields. While the thermal plasma emits X-rays through thermal bremsstrahlung and is easily detectable, the relativistic particles and magnetic fields, which range from  $\mathrm{\sim 0.1 - 10 \mu G}$, are only revealed through synchrotron emission from sources, known as diffuse radio sources, within the ICM \citep{Ferrari2008}. Such sources are categorized based on their location, size, and morphology within the cluster into radio halos, mini-halos, and radio relics. Radio halos are centrally located diffuse sources in merging clusters. They have roughly spherical structures with sizes of the order of megaparsecs (Mpc) and are co-spatial with the thermal intracluster medium (ICM). Mini-halos are also found near centers but are much smaller ($0.1-0.5$ Mpc) and are located in relaxed cool core clusters which also host a powerful radio galaxy associated with the BCG. Radio relics, on the other hand, have been defined as extended sources that show high levels of polarization ($10-50\%$  at GHz frequencies) and are located in the cluster periphery \citep{Van2019}. \\
The mechanism for generating radio halos is explained by the two most popular models. The first is turbulent re-acceleration or the primary model \citep{2007MNRAS.378..245B}, according to which, the electrons in the ICM are re-accelerated as a result of turbulence which is generated in the aftermath of a cluster merger. The second model is the hadronic or secondary model \citep{1980ApJ...239L..93D}, according to which, inelastic collisions between relativistic protons and thermal protons in the ICM produce pions which decay to produce electrons and gamma rays. Presently, turbulent re-acceleration is thought to be the main mechanism responsible for generating radio halos \citep{bru14}.
The theoretical models explaining the origin of relics suggest that they are the tracers of merger shocks \citep{ens98}. Relativistic particles are accelerated by
shocks, either by diffusive shock acceleration or adiabatic compression of fossil radio plasma \citep{ens02,hof04}. The diffuse radio emission in the ICM is also likely generated through the non-
gravitational feedback (heating) from the active galactic nuclei (AGN) found in the brightest central galaxy. Understanding the mechanism and properties of  non-thermal components of ICM is important for a comprehensive physical description of the intracluster medium in galaxy clusters, and plays a major role in the evolution of the large-scale structure of the Universe.\\
In this paper, we investigate the dynamical state and radio properties of the bimodal galaxy cluster Abell 1644 using low-frequency uGMRT band-2 (120–250 MHz) observations, complemented by Chandra X-ray data. We aim to characterize the radio emission associated with the brightest cluster galaxies (BCGs) and assess the presence of diffuse, nonthermal emission in the cluster environment. We further examine the X-ray surface brightness, temperature, and pseudo maps to identify signatures of gas sloshing and merger-related features, including cold fronts and temperature asymmetries. By combining the radio and X-ray results, we discuss the dynamical state of Abell 1644.

The paper is organised as follows: In sec 2, we give a brief description of our cluster Abell 1644. In sec 3, 
 we present the uGMRT Band-2 data analysis, followed by X-ray data analysis in sec 4. Results are presented in sec 5 followed by discussion in sec 6, and conclusion in sec 7. Throughout this work, we adopt a flat $\Lambda$CDM \
 \text{cosmology, using} H$_0$ = 70 kms$^{-1}$Mpc$^{-1}$, $\Omega_m = 0.3$, and $\Omega_\Lambda= 0.7$.
%\subsection{Subsection heading}

%\subsubsection{Subsubsection heading.} Subsubsection text goes here (Radhakrishnan {\em et al.} 1980).

\section{Abell 1644}
The galaxy cluster Abell 1644 (hereafter A1644) is a nearby galaxy cluster at  $ z = 0.0471 $\citep{tustin} located   $\sim 3$  Mpc from the Shapley super cluster \citep{filipis}. %It is known for its remarkable spiral-like X-ray emission.
X-ray studies have identified A1644 as a bimodal system \citep{jones}, comprising the sub-clusters; A1644S (main structure, on the South) and
A1644N (on the North). Both sub-clusters have their sharp X-ray peaks centred on a giant elliptical galaxy or their respective brightest cluster galaxies (BCGs). The main sub-cluster, A1644S, has a remarkable extended spiral-like X-ray morphology because of gas sloshing. Johnson et.al \cite{Johnson} attributed this gas sloshing in A1644S to the interaction with A1644N about 700 Myr ago or even earlier. More recently, a third sub-cluster (on the North) has been identified by Monteiro-Oliveira et.al \cite{olivier} from weak lensing analysis. Therefore, the previously known northern sub-cluster A1644N is renamed as A1644N1, and the new one as A1644N2. Although A1644N2 could not be seen on the X-ray images, its dynamical mass from the weak lensing model is not negligible. In fact, the weak lensing data suggest that northern sub-clusters have fairly the same masses,
$M_{200}^{N1} = 0.90^{+0.45}_{-0.85} \times 10^{14}$ and $M_{200}^{N2} = 0.76^{+0.37}_{-0.75} \times 10^{14} M_{\odot}$ ,
whereas the southern structure has mass $M_{200}^{S} = 1.90_{-1.28}^{+0.89} \times 10^{14} M_{\odot}$. Thus, A1644N2 may also have impacted the merging event in A1644. A numerical experiment by Doubrawa et.al \cite{Doubrawa} showed that the ICM slosh could have formed 1.6 Gyr ago as a result of the
interaction between A1644S and A1644N2.
Thus, for the spiral like X-ray emission,  the more likely scenario is a collision between A1644S and the newly discovered A1644N2,
where A1644N1 may be present as long as it does not greatly interfere in the formation of the spiral feature.\\
Radio studies of A1644 carried out in this work reveal two compact structures coinciding with the X-ray substructures. The third substructure revealed in weak lensing analysis is neither seen in X-ray nor in radio observation. Optical and near-infrared observations do not show any substructure within A1644 \citep{tustin}.
%\vspace{-2em}
\section{uGMRT band-2 observations and analysis}
The upgraded Giant Metrewave Radio Telescope (uGMRT) \rev{\citep{Gupta}} operates in four frequency bands. These bands are typically referred to as  Band-2, Band-3, Band-4, and Band-5, and they cover frequency ranges of  120-250 MHz, 300-500 MHz, 550-850 MHz, and 1050-1450 MHz, respectively. The uGMRT’s hybrid configuration, with baselines ranging from 100 m to 25 km, allows for high-resolution and sensitive imaging of extended sources across multiple frequencies. Observations of the complex galaxy cluster A1644 were conducted using uGMRT in band-2. Observations of this cluster were carried out on April 2, 2021 (PI Asif Iqbal) during night time to avoid radio frequency interference. Firstly, the amplitude calibrator, 3C147, was observed. Subsequently, the phase calibrator, 1311-222, was observed alternatively with the target observation. Finally, at the end of the observation, another amplitude calibrator, 3C286, was observed. The details of the observations for the cluster are given in Table \ref{table 1}. The data had 4096 channels, with each channel having a width of 48.828 kHz. \\
\begin{table}[h]
\centering
\caption{The uGMRT observation of A1644.}  
\label{table 1}
\begin{tabular}{ l  l  l  }
\hline
Frequency range & 120-250 MHz\\
Central frequency & 200 MHz  \\

%\hline
Bandwidth & 200 MHz \\
On source time& 7.8 hours  \\
Flux/Bandpass calibrators & 3C147, 3C286   \\
Phase calibrator & 1311-222 \\
Date of observation & 02-04-2021 
\\
\hline
\end{tabular}\\
\end{table}

\begin{figure}[h]
 \centering
\includegraphics[trim = 0.0cm 0.0cm 0.0cm 0.0cm, clip, height=10cm]{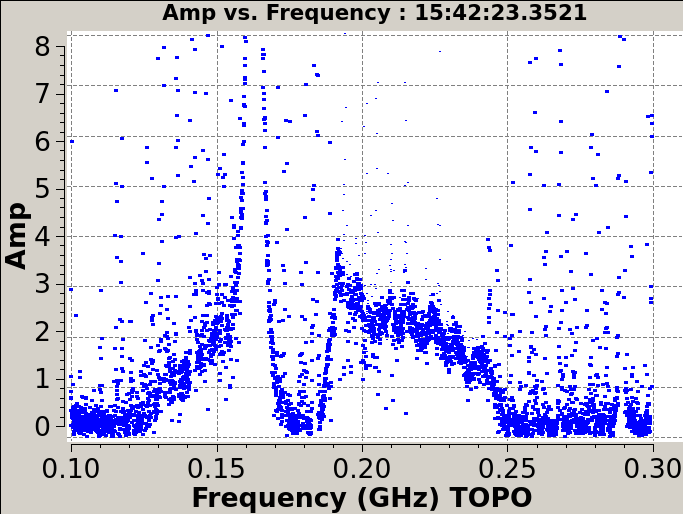}
    \caption{A representative plot of the band-2 spectrum on a baseline at a given time stamp. Baseline W03--E03 (`ll' correlation) was used to produce this plot. The amplitude is uncalibrated. The band has a notch filter around 0.17--0.19 GHz. The bright radio frequency interference in the frequencies below the notch filter, 0.15 - 0.17 GHz did not allow its usage.}
\label{fig:fig1}
\end{figure}

\begin{figure}[t]
 \centering
\includegraphics[width=0.9\textwidth]{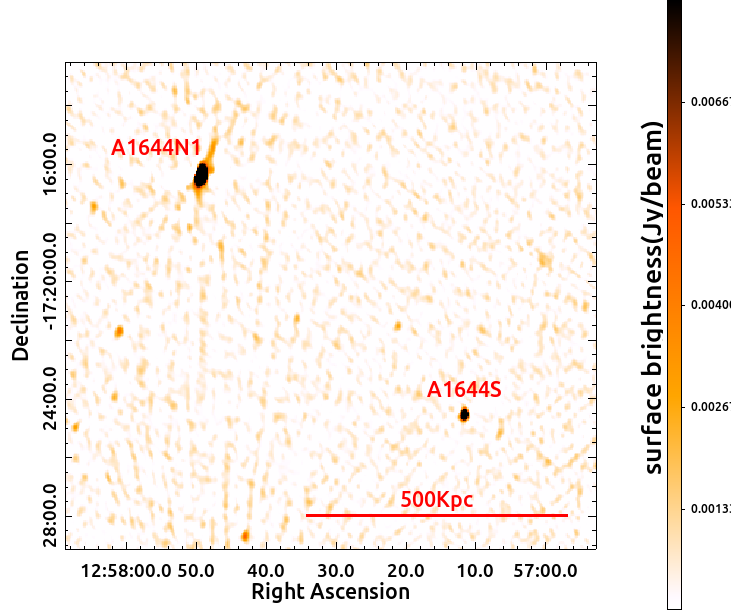}
    \caption{First uGMRT band 2 full resolution image of galaxy cluster A1644 obtained with robust=0.5. The rms is $\sigma_{rms} = 379$   $\ \mu Jybeam^{-1}$ and the restoring beam is $16.307'' \times 11.384'', P.A  -9.282\degree $.}
\label{fig:fig2}
\end{figure}

The data reduction process was performed using a continuum imaging pipeline called  CAsa Pipeline-cum-Toolkit for Upgraded Giant Metrewave Radio Telescope data REduction - CAPTURE \citep{rutacapture}. This pipeline is used for the first time in band-2 in this work, although it has been tested for band-3 and band-4 earlier. Data analysis of the cluster, A1644, was performed using a step-wise pipeline approach, where each step involved manual inspection of the data before proceeding to the next stage. Manual flagging was carried out wherever necessary to remove any unwanted data that the pipeline failed to remove. Prior to initiating the imaging process, calibration was performed several times to ensure proper calibration of the data. For the task \textit{tclean}, an interactive method was employed, and sources were manually masked to prevent overmasking, thereby generating an optimized mask. Self-calibration was then performed using the generated mask. At the end, one more round of \textit{tclean} was performed on the last visibility file using the \textit{auto-multithresh} technique to mask faint sources. This approach helped to restrict the generation of artifacts to the greatest extent possible. A significant portion of the data was lost because of high radio frequency interference (RFI) in band-2. After extensive RFI excision, approximately 75.6\% of the visibilities were flagged, leaving about 24\% of the data usable, corresponding to an effective bandwidth of $\approx 49$ MHz. As seen in Fig \ref{fig:fig1}, band-2 is RFI contaminated resulting in a significant data loss. Despite this data loss, the retained portion provided adequate sensitivity for reliable imaging, achieving an rms noise level of 379 µJy/Beam at a resolution of $16.307'' \times 11.384'', P.A  -9.282\degree $ , shown in  Fig.~\ref{fig:fig2}. 

\begin{figure}[t]
 \centering
\includegraphics[trim = 0.35cm 0.0cm 0.0cm 1.5cm, clip,height=6cm]{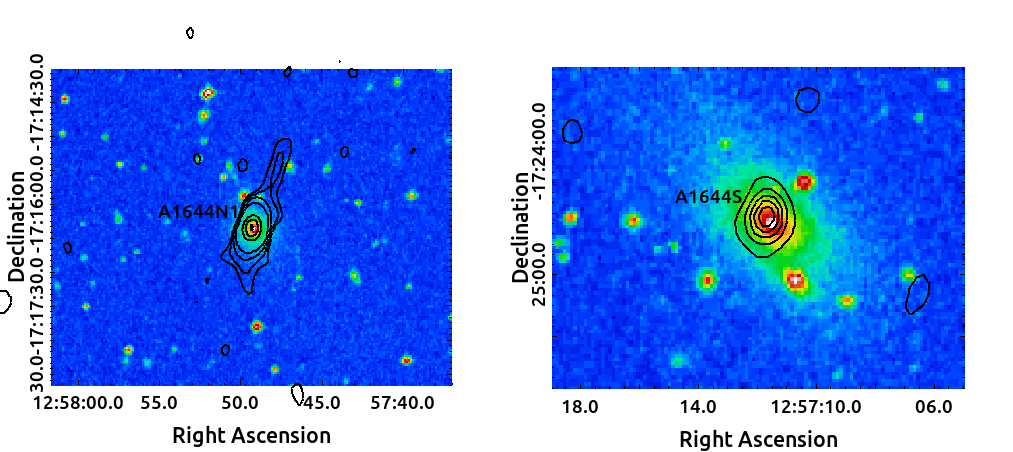}
    \caption{Digitized Sky Survey 2 (DSS2; \textcolor{red}{\citep{lasker}}) colour optical image of the Abell 1644 with overlaid radio continuum contours (Black).
Left: A1644N1; Right: A1644S. Black contours show the radio emission at 200 MHz. In both subclusters, the radio emission is compact and centered on the brightest cluster galaxy (BCG). Contours are drawn at $[1,2,4,8...] \times 3\sigma_{rms}$, where $\sigma_{rms} = 379 \ \mu Jybeam^{-1}$.}
\label{fig:fig3}
\end{figure}
\begin{figure}[t]
 \centering
    \includegraphics[width=0.9\textwidth]{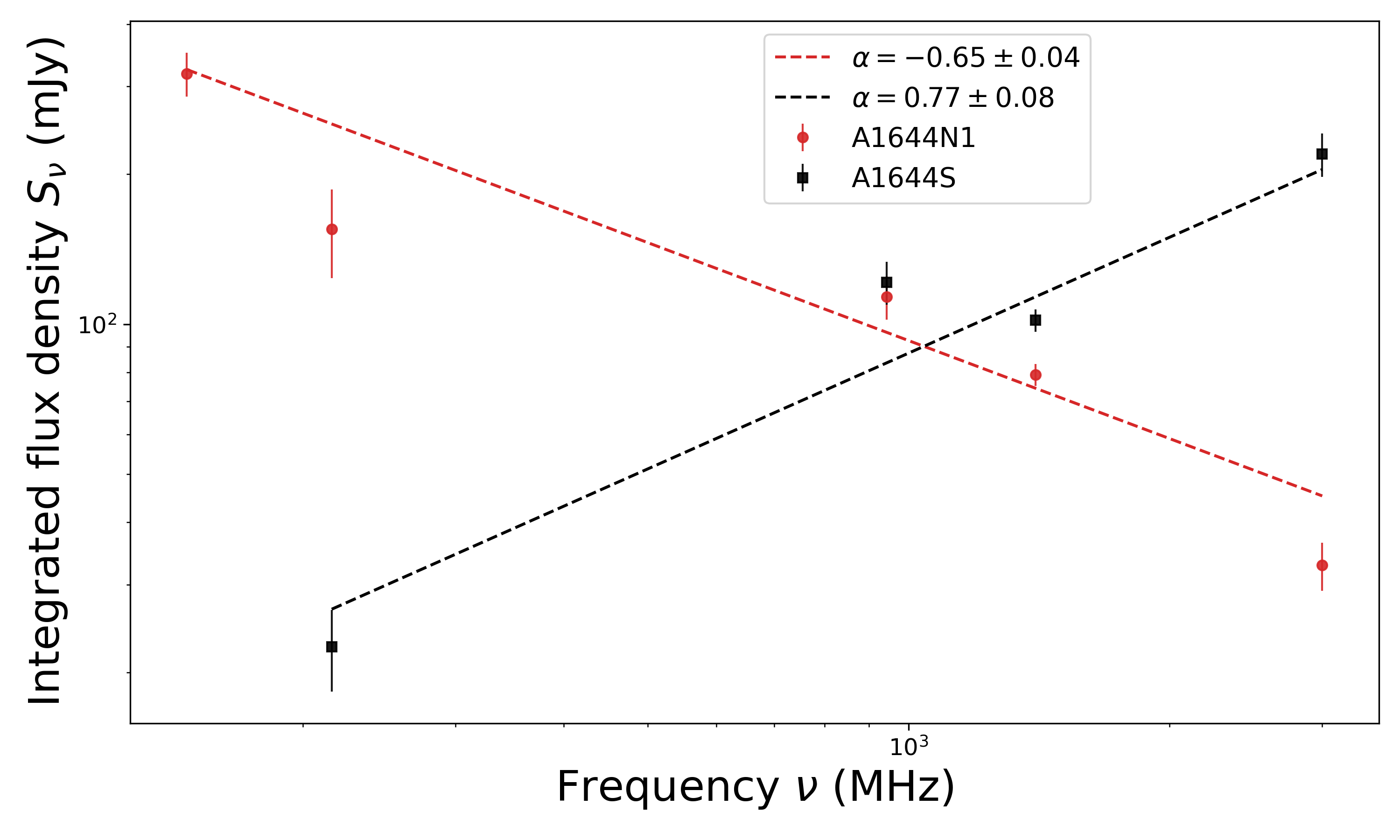} 
    \caption{Integrated radio spectrum of A1644N1 and A1644S. Error bars are the uncertainties in flux density. The dashed lines represents the best fit.}
\label{fig:fig4}
\end{figure}

In addition to the uGMRT observations, we used cutouts from the TGSS (TIFR GMRT Sky Survey \cite{tgss}), NVSS (NRAO VLA Sky Survey \cite{nrao}), \textcolor{red}{RACS (The Rapid ASKAP Continuum Survey \citep{Hotan2020})} and VLASS (VLA Sky Survey \cite{VLASS}) surveys at frequencies of 150 MHz, 943 MHz, 1.4 GHz, and 3 GHz, respectively, to obtain the flux density of two substructures.
All radio maps were convolved to have the same resolution of $45''$. The image was corrected for the
primary beam using the task ugmrtpb\footnote{https://github.com/ruta-k/uGMRTprimarybeam}
for the GMRT. We used CASA (Common Astronomy Software Applications; \citep{casa..127M})  for obtaining the flux densities of the two compact radio sources. \textcolor{red}{We measured the integrated flux density of A1644N1 and A1644S  using the \texttt{imview} tool within \texttt{CASA}. We manually selected the regions of emission above the $3\sigma$ level. The same regions were chosen for all maps (TGSS, band-2, RACS, NVSS, and VLASS) to maintain uniformity in the flux estimation}. The flux uncertainties were calculated by using the quadrature formula:  \begin{equation}
\delta S = \sqrt{(S_\nu \times \delta S_c)^2 + \textcolor{red}{\left(\sigma_{\rm rms}\sqrt{N_{\rm beam}}\right)^2}}
\end{equation}
where 
$\delta S$ is the uncertainty in flux density, $S_{\nu}$ is the measured flux density at frequency $\nu$, 
$\delta S_{\rm c}$ is the calibration error,  \textcolor{red}{$\sigma_{rms}$
 is the local rms noise of the image, and 
 $N_{\rm beam}$ is the number of synthesized beams contained within the selected source region.} We assumed  calibration uncertainty of 20\% for band-2 uGMRT data. \textcolor{red}{ For the survey data, we adopted calibration uncertainties of 10\% for TGSS \citep{tgss}, RACS \citep{askap} and VLASS \citep{lacy}, and 5\% for NVSS \citep{nrao}}. The flux density measurements are presented in Table~\ref{table:Table 2}, and the corresponding integrated radio spectrum is shown in Fig.~\ref{fig:fig4}. We define the spectral index $\alpha$ such that $S_{\nu} \propto \nu^{\alpha}$, where $S_{\nu}$ is the flux density at a given frequency $\nu$.\newline \textcolor{red}{Using the measured flux densities we then calculated the radio power at 200 MHz of these two central sources using the relation given in  \citep{Donoso}:
\begin{equation}
    P_{\nu} = 4\pi D_{L}^{2}\, S_{\nu}\,(1+z)^{-(1+\alpha)},
\end{equation}
where $P_{\nu}$ is the radio power, $D_{L}$ is the luminosity distance corresponding to the redshift $z$, $S_{\nu}$ is the measured flux density, and $(1+z)^{-(1+\alpha)}$ represents the standard $k$-correction term accounting for the spectral index $\alpha$. For A1644N1, power law fit to the integrated radio spectrum yeilded the best fit spectral index
$\alpha= -0.65$. For A1644S, the k-correction was computed using the spectral index  $\alpha=0.77$ obtained from a power-law fit to the higher-frequency data points (200 – 3000 MHz), excluding the TGSS measurement.}
\begin{table}

\centering

\caption{\textcolor{red}{Flux Density Estimates}}
\label{table:Table 2}
\subcaption{Flux Density of A1644N1}

\begin{tabular}{ l  l  l  }
\hline
Frequency(MHz)    &  Flux density(mJy) & Reference\\
\hline
\hline
$150$ & $318\pm32$ & TGSS\\
$200$& $155\pm31$ & This work\\
$943$& $113\pm11$ & RACS\\
$1400$& $79.1\pm4$ & NVSS \\
$3000$& $32.8\pm3$   & VLASS$-$QL\\
\hline

\end{tabular}

\bigskip
    
\subcaption{Flux Density of A1644S}  

\begin{tabular}{ l  l  l  }
\hline
Frequency(MHz)    &  Flux density(mJy) & Reference\\
\hline
\hline
$150$& $170\pm21$ & TGSS\\
$200$& $22\pm4$ & This work\\
$943$& $121\pm12$ & RACS\\
$1400$& $101\pm5$& NVSS \\
$3000$& $223\pm 21$ & VLASS$-$QL\\
\hline
\end{tabular}\\

\end{table}

\begin{table}
\centering
\caption{Results of the spectral fit along a 60\degree-wide sector (S2) encompasing the hot ICM region}
\label{table:Table 3}
%\subcaption{Flux Density of A1644N}
\begin{tabular}{ c c c c c c }
\hline
\hline
Bin no.    &  $\underset{(')}{R_{\mathrm{in}} - R_{\mathrm{out}}}$
&  $\underset{(\mathrm{keV})}{kT}$ & $\underset{(10^{-3}\,\mathrm{cm^{-3}})}{n_{\mathrm{e}}}$ & $\underset{(10^{-11}\,\mathrm{erg\,cm^{-3}})}{P}$ & Reduced $\chi^{2}$\\
\hline
1 & 0.1--0.4 &
$1.79\err{0.08}{0.09}$  &
$19.65\err{0.36}{0.37}$ &
$9.97\err{0.56}{0.51}$ &

1.12 \\

\multicolumn{6}{c}{\makebox[\textwidth][c]{\rule{0pt}{2.5ex}%
\leaders\hbox{--}\hfill\ \textit{Cold front}\ \leaders\hbox{--}\hfill}} \\

2 & 0.4--0.9 &
$4.28\err{0.74}{0.53}$ &
$6.20\err{0.12}{0.12}$ &
$7.65\err{1.33}{0.96}$ &

1.35 \\

3 & 0.9--1.3 &
$5.44\err{0.92}{0.68}$ &
$3.84\err{0.73}{0.75}$ &
$6.02\err{1.02}{0.76}$ &

1.33 \\

4 & 1.3--1.6 &
$10.24\err{3.44}{2.23}$ &
$3.35\err{0.07}{0.07}$ &
$9.89\err{3.82}{2.17}$ &

1.17 \\
\multicolumn{6}{c}{\makebox[\textwidth][c]{\rule{0pt}{2.5ex}%
\leaders\hbox{--}\hfill\ \textit{Possible shock}\ \leaders\hbox{--}\hfill}} \\

5 & 1.6--2.1 &
$5.20\err{0.74}{0.58}$ &
$2.65\err{0.04}{0.04}$ &
$3.98\err{0.57}{0.45}$ &

1.04 \\

6 & 2.1--3.0 &
$5.07\err{0.47}{0.41}$ &
$1.95\err{0.02}{0.02}$ &
$2.85\err{0.26}{0.23}$ &

1.02 \\
\\
\hline
\hline\\

\end{tabular}

\vspace{1mm}
{\footnotesize
\textbf{Notes.} Column 1: Radial bin number;
Column 2: inner and outer radius in arcmin ;
Column 3: temperature (KeV); Column 4: electron density; Column 5: pressure;
Column 6: reduced chi-square of the fit.
Errors correspond to the 68\% confidence level.}

\end{table}

% ALL FIGURES HERE!

%\end{wrapfigure}\\

\section{X-ray Analysis}
We analyzed archival  data for two separate observations [Obs-
IDs 7922 (PI Hudson; 2007), 2206 (PI Mohr; 2001)] of
A1644 having exposure times of 52 Ks and 20 Ks, respectively, with the Chandra X-ray observatory. The entire 72 Ks data were taken in VFAINT mode. We reduced the data using data reduction pipeline clusterPyXT \citep{Alden}. It is a new software pipeline that generates X-ray surface brightness, spectral temperature, density, and pressure maps from X-ray observations of galaxy clusters. It starts with the initialization of the cluster, for which the user has to provide key information:  name, metallicity, redshift, Chandra observation IDs (OBSIDs), and the galactic hydrogen column density ($N_{H}$). For spectral fitting, we adopted  $N_{H}=5.69 \times 10^{20} cm^{-2}$, and the metallicity of 0.50 solar \citep{accept}. We ensured a signal-to-noise ratio of 50 to optimize the accuracy of our spectral fits while minimizing background contamination.

After inputting the required  information, the pipeline automatically downloads data from the Chandra archive using the CIAO (Chandra Interactive Analysis of Observations)
task $download\_chandra\_obsid$, and cleans it in a standard manner
for both data and background using task $chandra\_repro$. Following this, it merges all ObsIds to create a combined flux map. Next, point sources need to be removed from the flux map (X-ray surface brightness image). We used \textit{wavedect} task to detect point sources in the 0.2–12 keV band with the scales of 1, 2, 4, 8, and 16 pixels and created a region file containing location of all point sources that are not part of cluster. We also checked visually for any false detection or if \textit{wavedect}  failed to detect any real point source. The pipeline then creates a new flux image without point sources. This is followed by filtering out time intervals containing high energy background flares from each observation. The corresponding high-energy flares are removed by generating light curves in the full band and 9.5–12 keV band. The light curves are binned at 259 seconds per bin. Count rates greater than $3\sigma$ from mean are removed using the task  \textit{deflare}. We visually examined the light curve to ensure that the background flares were effectively removed.

\subsection{X-ray surface brightness map}
\begin{figure*}[htbp]
    \centering
 %   \begin{minipage}{0.45\linewidth}
    \includegraphics[trim = 0.3cm 0.0cm 0.2cm 2.0cm, clip,height=7cm]{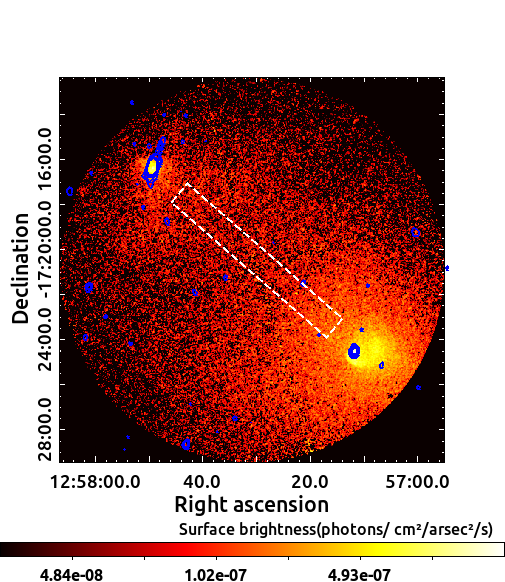}     
   % 
%\end{minipage}
%\hfill
%\begin{minipage}{0.45\linewidth}
%\centering
 \includegraphics[trim = 0.3cm 0.0cm 0.2cm 2.0cm, clip,height=7cm]{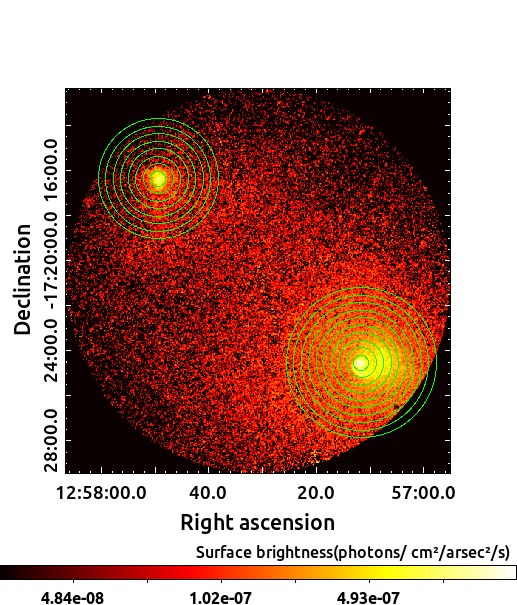}
% \end{minipage}
 \caption{Exposure corrected, background subtracted, point source removed,
surface brightness map of galaxy cluster A1644 in 0.7-8.0 keV energy range. \emph{Left}: Radio contour levels (blue) drawn at $[1,2,4,8...] \times 3\sigma_{rms}$, where $\sigma_{rms} = 379 \ \mu Jybeam^{-1}$ are overlaid. \textcolor{red}{The white dashed rectangle is the bridge region used to find temperature between the two substructures.}  \emph{Right}: Green annular regions on the right image are drawn to get the temperature profile of the two substructures of the galaxy cluster. } 
\label{fig5}
 \end{figure*}

The exposure corrected, background and point sources subtracted, 0.7-8.0 keV surface brightness image is shown in  Fig.~\ref{fig5}.
We have used the CIAO task \textit{dmfilth} to fill in the holes of the removed point sources. However, because of the low count rate in fluximage, \textit{dmfilth} worked only after we first scaled up fluximage using task \textit{dmimgcalc}. We then rescaled it back after filling up point source regions.
\begin{figure}
\begin{minipage}{0.5\linewidth}
    \centering
    \includegraphics[width=0.9\textwidth]{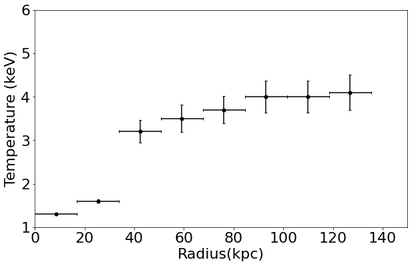}

\label{fig:sb}
 \end{minipage}
\hfill
\begin{minipage}{0.5\linewidth}
    \centering
    \includegraphics[width=0.9\textwidth]{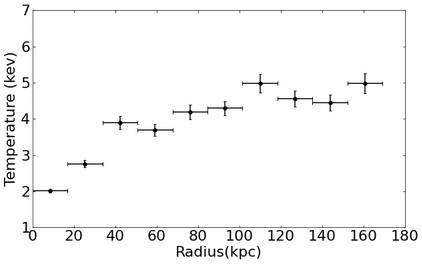}

 \end{minipage} 
 \caption{ Temperature profiles corresponding to the annuli shown in Fig. \ref{fig5}. \emph{Left}: Temperature profile of A1644N. \emph{Right}: Temperature profile of A1644S.\newline Error bars indicate the $1\sigma $statistical uncertainties on the APEC temperature obtained from Sherpa fits using the chi2xspecvar statistic. Horizontal bars represent the width of the annuli.}
\label{fig:fig6}

\end{figure}

 \subsection{Temperature map}

%\begin{wrapfigure}{R}{0.5\textwidth}
 To generate temperature maps, ClusterPyXT uses the Astrophysical Plasma Emission Code (APEC) model for optically thin collisionally ionized hot plasma combined with a photoelectric absorption (PHABS) model \citep{Smith,Balucinska-Church}. These models include important factors like redshift, metallicity, temperature, normalization and Hydrogen column density. The energy and counts of each photon are fit to the data using temperature and normalization as the free parameters. ClusterPyXT \citep{Alden} uses an adaptive circular binning ACB \citep{datta} method to create a bin for each pixel. The bin size is determined on the size of the circle needed to achieve a desired signal-to-noise ratio. This method creates a bin for every pixel, producing a high-resolution temperature map, shown in Fig.~\ref{fig:fig7} and the entire process is automated in ClusterPyXT.
 We also get pseudo pressure and density maps as a byproduct in clusterPyXt, also shown in Fig.~\ref{fig:fig7}.

\subsection{Spectral Extraction and Fitting}
\begin{figure}[t]
   \centering
%   \begin{minipage}{0.45\linewidth}
  % \centering
    \includegraphics[width=0.9\textwidth]{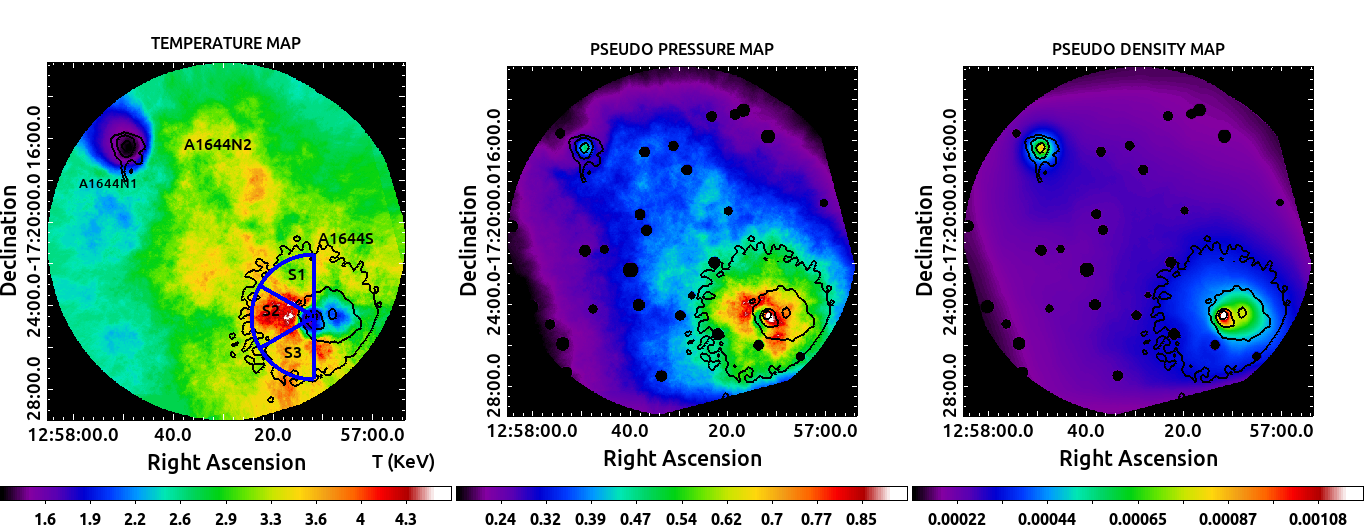} 
    \includegraphics[width=0.9\textwidth]{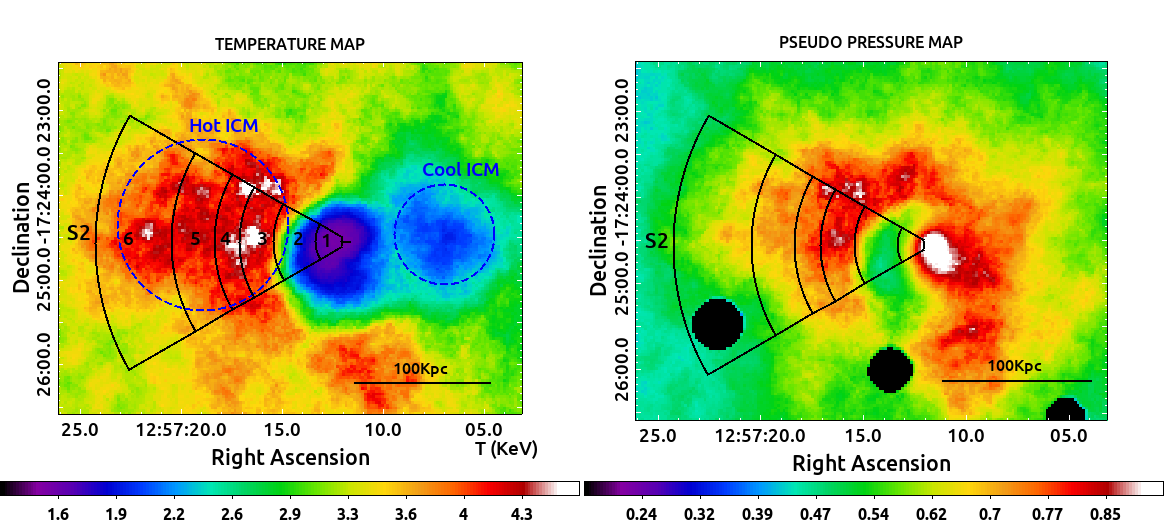}
   
\caption{\emph{Top:} X-ray surface brightness contours (black) overlaid on the thermodynamic maps: temperature (left), pseudo-pressure (middle), and pseudo-density (right). The temperature map also shows the three azimuthal sectors used to find surface brightness discontinuity, namely S1 (90°–150°), S2 (150°–210°), and S3 (210°–270°). The locations of the three main substructures identified from weak-lensing analysis are marked on the temperature map. Black circles in the pseudo-pressure and pseudo-density maps mark removed point sources. 
\emph{Bottom:} Zoomed-in views of the temperature and  pseudo-pressure maps showing the wedge-shaped 
annuli, S2, divided into six radial bins ( Bin 1-6) to extract the temperature profile. The blue dashed circles highlight the regions of hot and cold ICM identified in the temperature map. 
}

\label{fig:fig7}

\end{figure}

\textcolor{red}{We extracted X-ray spectra using the CIAO task specextract. For the radial temperature profiles of the two substructures, A1644N1 and A1644S, spectra were extracted from a set of concentric circular annuli centered on their respective X-ray surface brightness peaks. Specifically, the annuli for A1644N1 are centered at (12:57:49.3124, -17:16:23.379), while those of A1644S are centered at (12:57:11.6505, -17:24:33.389). The corresponding radial temperature profiles derived from these annuli are shown in Fig. \ref{fig:fig6}. In addition, to study hot ICM region east of A1644S}, spectra were extracted from wedge-shaped annuli spanning a position angle range of $150^\circ$--$210^\circ$( labelled S2), centered on the X-ray brightness peak of A1644S. These wedge annuli extend out to a radius of 3 arcmin and were divided into six radial bins.
Spectral modeling was performed in Sherpa over the 0.5–5.0 keV energy range. We adopted a single-temperature thermal plasma model of the form phabs $\times$ apec, where the absorbing column density was fixed to the Galactic value $N_{H}=5.69 \times 10^{20} cm^{-2}$. The redshift was fixed at z=0.047, and the metal abundance was held constant at 0.5 solar. Photon grouping was applied to ensure a minimum of 20 counts per bin, and fits were evaluated using the $\chi^2$statistics with variable weighting (chi2xspecvar). For each spectrum, the APEC temperature and normalization were allowed to vary while all other parameters were frozen.The APEC normalization yields the emission integral,
$\mathrm{EI} = \int n_{\mathrm{e}}\, n_{\mathrm{p}}\, dV$
through the relation
\begin{equation*}
    \mathrm{norm} = \frac{10^{-14}}{4\pi D_{\mathrm{A}}^{2} (1+z)^{2}} \, \mathrm{EI}
\end{equation*}
where \(D_{\mathrm{A}}\) is the angular diameter distance. Assuming a fully ionized
plasma with \(n_{\mathrm{e}} = 1.2\, n_{\mathrm{p}}\), the measured APEC normalizations were
used to determine the electron density in each annulus. The thermal pressure was then computed as $P =(n_{\mathrm{e}} + n_{\mathrm{p}})$ kT. 
The best fit parameters obtained from the fits to the annuli spectra and the derived electron density and pressure are summarized in table \ref{table:Table 3}.
\textcolor{red}{The temperature profile along the S2 sector shows a significant  increase in bin~4, where the best-fit temperature reaches 
$kT = 10.24^{+3.44}_{-2.23}\,\mathrm{keV}$, compared to $kT \approx 5.5\,\mathrm{keV}$ in bin~3. However, the projected temperature map ( Fig. \ref{fig:fig7}) 
is not apparently  consistent with the temperature of bin 4 extracted from spectral fitting. 
The difference can be due to: 1) the temperatures in Table 3 are derived from emission-weighted spectral fits, whereas the 2D map is based on adaptively binned projected regions optimized for visualization. These different spatial weighting and binning approaches can produce moderate discrepancies; 2) Localized shock-heated or thermally complex gas may elevate the emission-weighted temperature of the integrated spectrum without dominating the projected 2D map visually; and 3) Bin 4 has relatively large uncertainties. When the quoted errors are properly taken into account, the temperatures are consistent with the 2D map within $2\sigma$.}

\textcolor{red}{Besides, we tried to see if there are any unresolved point sources that are contributing to this discrepancy by checking the point source removed event file. We reassured that there are no such point sources and therefore the possibility of mismatch due to unresolved point sources is ruled out.}

%\subsection{\tcr{Density maps}}

%\tcr{I think you should discuss Norm maps instead od density maps - they are equivalent to the density and are of higher resolution- the density maps discussion seems to be irrelevant here}
 %To analyze the spatial distribution of the intracluster medium (ICM) within the galaxy cluster, we examined the electron density map generated as a byproduct of the ClusterPyXT pipeline \citep{Alden} %\tcr{why you cite this paper here: it looks pyxt is developed by eckert}.
% Density maps could provide crucial insight into the structure and composition of the cluster, particularly in the core and outskirts. By visualizing the density variations across the cluster environment, we can gain a deeper understanding of its morphological features, dynamical state, and potential substructures.
%\begin{figure}
%\begin{minipage}{0.45\linewidth}
% \flushleft
%\includegraphics[trim = 0.0cm 0.0cm 0.0 cm 2cm, clip,height=8.5cm]{density_map_cropped_with_sectors.png} 
%\caption{Electron density map of the A1644 intracluster medium (ICM) derived using the ClusterPyXT pipeline. The color scale represents the electron density, \( n_e \), in units of cm\(^{-3}\), ranging from \(1.6 \times 10^{-4}\) to \(11.5 \times 10^{-4}\).}

%\label{fig:desnity_map}
%\end{minipage}

%\end{figure}

\section{Results}

\begin{figure*}[h]
\begin{minipage}{0.5\linewidth}
    \centering
    \includegraphics[width=0.9\textwidth]{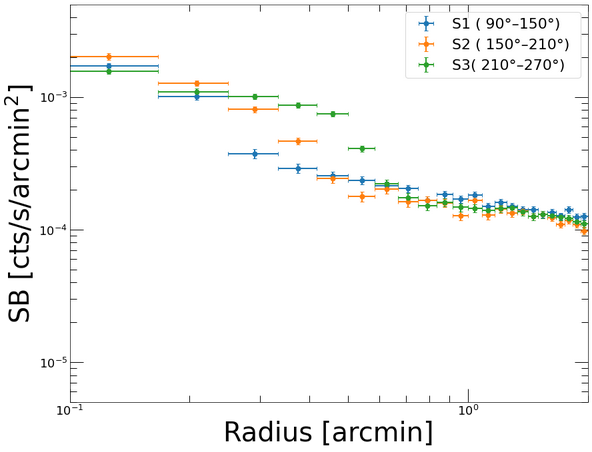}
\label{fig:sb}
 \end{minipage}
\hfill
\begin{minipage}{0.5\linewidth}
    \centering
    \includegraphics[width=0.9\textwidth]{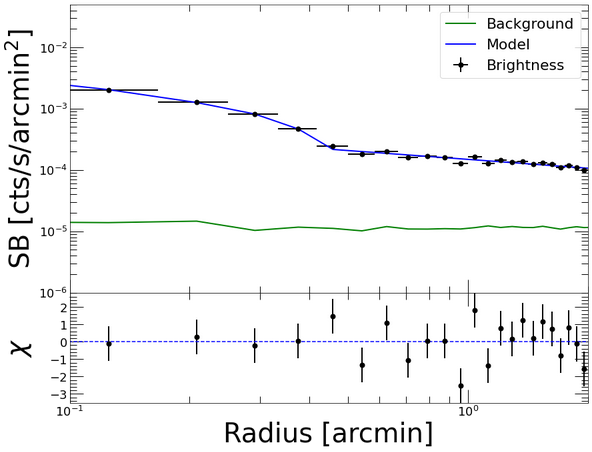}

 \end{minipage} 
  \caption{\emph{Left}: Surface brightness (SB) profiles of A1644S extracted in 5$^{\prime\prime}$-wide annular bins within three 60$^{\circ}$-wide sectors covering position angles (90$^{\circ}$–270$^{\circ}$). 
 The surface brightness discontinuity shifts outward from $\sim$0.3~arcmin in the S1 to $\sim$0.5~arcmin in the S3.
 \emph{Right}: Surface brightness profile within S2. As mentioned in text, the discontinuity is around 0.4 arcmin with a density jump of 2.79.  The best-fit broken power-law model is overlaid in blue, while the residuals are shown in the lower box. Radial distances are shown in units of arcminutes. }
\label{fig:fig8}

\end{figure*}

In Fig.~\ref{fig:fig2}, we show the uGMRT band-2 full resolution continuum image of the galaxy cluster A1644. We detected two compact radio sources corresponding to the X-ray peaks of the respective substructures: A1644N1 and A1644S. These radio sources are associated with the brightest cluster galaxies (BCGs) at the centers of the northern and southern substructures, as seen in  Fig.\ref{fig:fig3}, indicating that they are likely powered by active galactic nuclei (AGN).\\ \newline
In Fig.~\ref{fig:fig4}, we show the integrated flux densities of A1644N1 and A1644S at five frequencies: 150, 200, 943 1400, and 3000 MHz, using cutouts from TGSS, our uGMRT Band-2 image, RACS, NVSS, and VLASS respectively.
For A1644N1, the flux density follows a well-defined power-law trend in log–log space, with a best-fit spectral index of $\alpha = -0.65 \pm 0.04 $, consistent with synchrotron emission from AGN. The corresponding radio power at 200 MHz is $P_{A1644N1} = 7.9\times 10^{23} $ W/Hz.
In contrast, A1644S shows a non-monotonic spectral behaviour: the flux density decreases sharply from 150 MHz to 200 MHz and then increases at 943 MHZ, 1.4 GHz and 3 GHz. A power law fit to higher frequncy points (excluding TGSS) yeilds a spectral index of $\alpha= 0.77 \pm 0.08$ , and the estimated radio power at 200 MHz is $1.0\times 10^{23}$ W/Hz. This inverted spectrum suggests the presence of synchrotron self-absorption at low frequencies, likely
originating from a compact AGN core embedded within the southern
subcluster.\\ \newline
In the left panel of  Fig.~\ref{fig5}, we present the surface brightness map of our cluster with overlaid radio contours drawn at $[1,2,4,8...]\times 3\sigma_{rms}$, where $\sigma_{rms} = 379.0 \ \mu Jybeam^{-1}$ and on right panel we have drawn annular regions to obtain the temperature profile of each substruture. The annuli of both substructures are centered at respective X-ray emission peaks. The associated radial temperature profiles for these annular regions are shown in Fig. \ref{fig:fig6}. We see that both substructures are cool-core with a dip in the temperature profiles towards the cluster center. We determined the average temperature within a 0-300 Kpc region for each substructure. The average temperature for A1644N1 was found to be $T_{avg, A1644N1}=2.84 \pm 0.11 keV$, while for A1644S it was $T_{avg, A1644S}= 3.93  \pm  0.84keV$.\\ \newline
In Fig~\ref{fig:fig7}, we present the temperature, pseudo-pressure, and pseudo-density maps of A1644 in the top panel. The bottom panel shows a zoomed-in view of the temperature and pseudo-pressure maps, highlighting the region of interest from which the wedge annuli were extracted. The temperature map reveals several previously unexplored features in this cluster. In particular, we detect a hot ICM region to the east of A1644S, as well as a colder structure to the west of it. To better understand the structures observed in the temperature map, we first derived the surface brightness profile of A1644S. This profile shows a clear discontinuity in the eastern direction. To characterize this feature, we extracted surface brightness profiles in three azimuthal sectors toward the east (labeled S1, S2, and S3), using circular sectors with varying opening angles. The resulting profiles were fitted using pyproffit \citep{eckert} with a broken power-law density model.
A prominent surface brightness edge is detected in all three sectors. In S1, the edge is relatively weak and occurs at a radius of $\approx 0.3$ arcmin, with a density jump of $\approx 2.3$. In S2, the edge shifts outward to 0.406 arcmin with a density jump of 2.79. The edge becomes more prominent in S3, where it shifts to approximately 0.5 arcmin with a density jump of around 3.2 (Fig. \ref{fig:fig8}). No surface brightness discontinuity is detected in the outer region associated with the hot structure.\\
To further investigate the nature of the hot ICM region, we performed a spectral analysis of the S2 sector, which encompasses the hot region. 
We extracted X-ray spectra of the ICM within S2 (shown in Fig.~\ref{fig:fig7}), covering radii up to 3~arcmin and divided into six spectral bins. Each spectrum was fitted in the 0.5-5.0~keV energy band using an APEC model to derive the gas temperature and electron density. These measurements were then combined to estimate the pressure profile, and the results are summarized in Table~\ref{table:Table 3}. Within this sector, we identify an inner edge located at 0.406~arcmin (between bins~1 and~2) from the cluster X-ray peak, characterized by a density jump of $J = 2.79$.

Across this edge we measured a temperature jump of 
$kT_{\mathrm{in}} / kT_{\mathrm{out}} = 0.42$ and a pressure ratio of 
$P_{\mathrm{in}} / P_{\mathrm{out}} = 1.1$.
The nearly constant pressure and the lower temperature inside the edge are consistent with the interpretation of this feature as a \textit{cold front}, which is in agreement with the previous results  \citep{Johnson}.
However, between bin 4 and bin 5, the spectral analysis revealed a sharp increase in temperature and pressure, with 
$kT_{\mathrm{in}} / kT_{\mathrm{out}} = 1.97$ and 
$P_{\mathrm{in}} / P_{\mathrm{out}} = 2.5$.
These values are consistent with the presence of a possible shock front, which would have heated and compressed the ICM following its passage. However, no corresponding surface brightness edge was detected at this location, nor there is a boundary visible in the temperature map which puts a question mark on the definite presence of a shock front.

\begin{figure}[tbp]
\centering
    \includegraphics[width=0.5\textwidth]{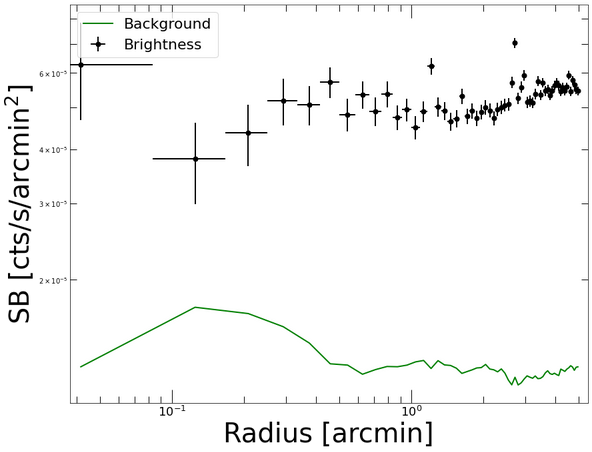}

\caption{Surface brightness profile of the region between the two substructures; A1644S and A1644N.}    
\label{fig:fig9}

 \end{figure}

Finally, we examined the X-ray emission between the two subclusters. The projected separation between the X-ray peaks of A1644N1 and A1644S is approximately 674 kpc, consistent with the 700 kpc separation reported by Monteiro-Oliveira et al. \citep{olivier}. To estimate the temperature of the inter-cluster gas, we defined the bridge region in such a way so that the central cool-core regions of
both subclusters are excluded  (fig \ref{fig5}). The region begins at the radius where the temperature profile of both subclusters
becomes approximately uniform, ensuring that the selected bridge area primarily traces the inter-cluster gas and minimizes contamination from the inner subcluster emission.  A spectral fit to this bridge region yielded a best-fit temperature of $kT = 5.3$ keV. To assess the reliability of this temperature measurement, we extracted the surface brightness profile between the two subclusters using pyproffit. The resulting profile is shown in Fig. \ref{fig:fig9}

\section{Discussion}

The present work provides a joint radio and X-ray study of the merging galaxy cluster Abell 1644 using uGMRT Band--2 (125--250 MHz) observations and archival \textit{Chandra} data. As a merging system reported in the literature, Abell 1644 provides an ideal laboratory for multiwavelength studies. We therefore analysed radio observations to search for diffuse emission and combined them with X-ray data to examine the thermodynamic properties of the ICM.

\subsection{Radio morphology and absence of diffuse emission}

Our Band--2 uGMRT image reveals only two compact radio sources that spatially coincide with the X-ray peaks of the respective substructures (A1644N1 and A1644S). No evidence of diffuse emission such as radio halos or relics, is  detected within the sensitivity limits of our data. The absence of diffuse radio features is noteworthy given that Abell 1644 has been classified as a merging cluster in previous studies \citep[e.g.][]{ Johnson, olivier}. 
In such systems, turbulence induced by the interaction is expected to reaccelerate relativistic electrons, producing extended radio halos \citep{Van2019}. 
The lack of such emission in A1644 may suggest that the merger was relatively mild, not injecting enough turbulence for large-scale reacceleration or the system is in a later post-merger phase, when turbulence has decayed, and relativistic electrons have cooled.

\subsection{Thermodynamic properties of ICM}

Complementing the radio observations, our X-ray analysis reveals a cold front located east of the X-ray peak of A1644S. This feature corresponds to the sloshing front previously reported by Johnson et al. \citep{Johnson}, and is attributed  to an interaction between A1644S and A1644N1. However, recent dynamical work by Monteiro-Oliveira et al. \citep{olivier} revises this picture, showing that the sloshing in A1644S was instead triggered by the passage of A1644N2. Our detection is therefore consistent with this updated interpretation, indicating that the cold front is a remnant of the past encounter between A1644S and A1644N2.
The temperature map further supports this scenario by revealing a pronounced asymmetry in the intracluster medium. The cool core of A1644S is flanked by a hotter ICM region to the east and a cooler region to the west, a characteristic signature of gas sloshing induced by a merger. Weak-lensing results and tailored hydrodynamical simulations indicate that this sloshing was driven primarily by a close, off-axis encounter with the gas-poor subcluster A1644N2. During its pericentric passage approximately 1.6 Gyr ago, A1644N2 was stripped of most of its gas while its gravitational perturbation displaced the cool core of A1644S.  \textcolor{red}{The resulting oscillatory motion displaced cold, low-entropy gas, producing a spiral cold front to the east of the core, while a hotter ICM region is observed at larger radii in the same direction.} The observed temperature distribution thus captures the  dynamical aftermath of this interaction and highlights the significant role of gas-poor subclusters in shaping the thermal structure of merging galaxy clusters.\\ 
To further study the hot ICM region, we performed spectral extraction across this region. We tested multiple extraction geometries and found that the thermodynamic discontinuities become apparent only for the sector labeled S2. In this sector, we detect a jump in temperature and pressure at a radius of approximately 1.6 arcmin, which is suggestive of a possible shock. However, the absence of a corresponding surface-brightness edge and the lack of a clearly defined boundary in the temperature map at this location does not favour classification as a shock with certainty. \\
Overall, our findings reinforce the interpretation of Abell 1644 as a post-merger, sloshing system with possible weak shocks but no evidence of large-scale nonthermal diffuse emission.
Future deep radio observations could help constrain the presence of any faint diffuse emission below our current sensitivity limits. Likewise, deeper Chandra or XMM-Newton exposures could confirm the temperature and surface brightness jumps, and better constrain the Mach number and energetics of the possible shock.

\section{Summary and conclusions}
We have used two complementary probes (radio and X-ray) to study the merging galaxy cluster Abell 1644. The X-ray analysis confirms the presence of a cold front in the southern subcluster (A1644S), reported previously \citep{Johnson}. 
 We also identify a temperature asymmetry, with a hot ICM region to the east of A1644S and a cooler ICM region to its west. This temperature asymmetry may possibly be a remnant of the past merger as it is not associated with the major shock front. No diffuse radio emission is detected, suggesting that merger driven turbulence has largely dissipated or remains below the threshold required for particle reacceleration.
 Taken together, these results point to a scenario in which Abell 1644 is a post-merger system that retains evidence of its dynamical past through sloshing and possible residual shocks, but which has evolved beyond the turbulent phase that typically powers cluster-scale synchrotron emission. This post-merger scenario is consistent with the picture in which A1644S and A1644N2 subclusters have already undergone a core passage, 
reached apocentric separation, and are now moving towards a possible second encounter \textcolor{red}{\citep{olivier}}. 
This combination of a preserved cool core, a sloshing front, and the absence of diffuse radio emission makes A1644 a valuable case for studying the late stages of cluster mergers and the thermalization of merger energy in the ICM.

\acknowledgments
This work is supported by the Science and Engineering Research Board (SERB), Department of Science and Technology, Government of India, under CRG/2021/002685. 
R.K. acknowledges the support of the Department of Atomic Energy, Government of India, under project no. 12-R\&D-TFR-5.02-0700 and from the SERB Women Excellence Award WEA/2021/000008. We thank the staff of the GMRT who made these observations possible. The scientific results discussed in this paper are partly
based on data obtained from the Chandra Data
Archive provided by Chandra X-ray Center (CXC).
We acknowledge the use of the TIFR GMRT Sky Survey (TGSS). We acknowledge the use of the VLA Sky Survey (VLASS), a scientific product of the National Radio Astronomy Observatory (NRAO). We acknowledge the use of  The Digitized Sky Survey produced at the Space Telescope Science Institute under U.S. Government grant NAG W-2166.
We acknowledge the use of NASA/IPAC Extragalactic
Database (NED), operated by the JPL, Caltech, under contract with NASA.\\
\emph{Facilities}: upgraded Giant Metrewave Radio Telescope (uGMRT), Chandra\\
\emph{Software}: CASA, matplotlib, ds9
\vspace{-1em}

% Bibliography

%% [A] Recommended: using JHEP.bst file
%% \bibliographystyle{JHEP}
%% \bibliography{biblio.bib}

%% or
%% [B] Manual formatting (see below)
%% (i) We suggest to always provide author, title and journal data or doi:
%% in short all the informations that clearly identify a document.
%% (ii) please avoid comments such as "For a review'', "For some examples",
%% "and references therein" or move them in the text. In general, please leave only references in the bibliography and move all
%% accessory text in footnotes.
%% (iii) Also, please have only one work for each \bibitem.

\bibliographystyle{JHEP}
\bibliography{A1644}
\end{document}